\documentclass[letterpaper,journal]{IEEEtran}

\usepackage{amsmath,amssymb,amsfonts}
\usepackage{graphicx}
\usepackage{cite}
\usepackage{url}
\usepackage{textcomp}
\usepackage[hidelinks]{hyperref}

\begin{document}

\title{Control of Harmful Information Spreading on Adaptive Higher-Order Networks via Group Dissolution}

\author{Longzhao~Liu, Zhihao~Han, Xingru~Chen, Chunyu~Luo, Hongwei~Zheng, Xin~Wang, and Shaoting~Tang%
\thanks{This work is supported by Program of National Natural Science Foundation of China (12425114, 12201026, 12501702, 12501718, 62441617), the Fundamental Research Funds for the Central Universities, Beijing Natural Science Foundation (Z230001), and Beijing Advanced Innovation Center for Future Blockchain and Privacy Computing. \emph{(Longzhao Liu and Zhihao Han contributed equally to this work.)} \emph{(Corresponding authors: Xin Wang and Shaoting Tang.)}}%
\thanks{Longzhao Liu and Xin Wang are with the Institute of Artificial Intelligence, Beihang University, Beijing 100191, China, also with the Key Laboratory of Mathematics, Informatics and Behavioral Semantics, Beihang University, Beijing 100191, China, also with the Beijing Advanced Innovation Center for Future Blockchain and Privacy Computing, Beihang University, Beijing 100191, China, and also with the State Key Laboratory of Complex \& Critical Software Environment, Beihang University, Beijing 100191, China (e-mail: \mbox{longzhao@buaa.edu.cn}; \mbox{wangxin\_1993@buaa.edu.cn}).}%
\thanks{Zhihao Han is with the Institute of Artificial Intelligence, Beihang University, Beijing 100191, China, and also with the Key Laboratory of Mathematics, Informatics and Behavioral Semantics, Beihang University, Beijing 100191, China (e-mail: \mbox{hanzhihao@buaa.edu.cn}).}%
\thanks{Xingru Chen and Chunyu Luo are with the Institute of Artificial Intelligence, Beihang University, Beijing 100191, China (e-mail: \mbox{xingrucz@gmail.com}; \mbox{22377175@buaa.edu.cn}).}%
\thanks{Hongwei Zheng is with the Beijing Academy of Blockchain and Edge Computing, Beijing 100085, China (e-mail: \mbox{hwzheng@pku.edu.cn}).}%
\thanks{Shaoting Tang is with the Institute of Artificial Intelligence, Beihang University, Beijing 100191, China, also with the Hangzhou International Innovation Institute, Beihang University, Hangzhou 311115, China, also with the Key Laboratory of Mathematics, Informatics and Behavioral Semantics, Beihang University, Beijing 100191, China, also with the Institute of Medical Artificial Intelligence, Binzhou Medical University, Yantai 264003, China, also with the Beijing Advanced Innovation Center for Future Blockchain and Privacy Computing, Beihang University, Beijing 100191, China, and also with the State Key Laboratory of Complex \& Critical Software Environment, Beihang University, Beijing 100191, China (e-mail: \mbox{tangshaoting@buaa.edu.cn}).}}

\maketitle

\begin{abstract}
Curbing harmful information contagion remains a critical challenge, motivating platform-level interventions such as group dissolution to sever transmission chains. However, in practice, users affected by dissolution often exhibit adaptive behavior, rewiring to form new groups. Yet, it remains unclear how these two mechanisms jointly shape information contagion and whether group dissolution remains effective in suppressing it. Here, we develop an adaptive higher-order contagion model that integrates platform-induced group dissolution with user adaptive rewiring, and derive a theoretical framework. Notably, we reveal an effective window for group dissolution, bounded by a critical infection rate. Above this threshold, dissolution backfires and amplifies information prevalence. Within this window, dissolution acts non-monotonically, initially exacerbating prevalence before eradicating contagion via a discontinuous transition beyond a critical dissolution rate. We further show that higher-order reinforcement expands this infection-rate window over which dissolution remains effective, whereas rewiring homophily substantially narrows it. Simulations on empirical hypergraph also validate these findings. Our work highlights the interplay between top-down platform interventions and bottom-up user adaptation, underscoring the need to account for adaptive responses when designing strategies to curb harmful information without unintended amplification. Our code is available at
\url{https://github.com/hzhbuaa/Adaptive-Higher-Order-Contagion-of-Harmful-Information}.
\end{abstract}

\begin{IEEEkeywords}
Adaptive hypergraphs, contagion dynamics, harmful information, platform interventions.
\end{IEEEkeywords}

\section{Introduction}\label{sec1}

Harmful information, such as misinformation, online hate speech and extremism, poses a severe threat to society \cite{lazer2018science, loomba2021measuring, muller2021fanning, johnson2016new}. This issue has been further amplified by the development of social media platforms such as X (Twitter) and Facebook, which significantly accelerate the spread of such content \cite{del2016spreading, vosoughi2018spread, cinelli2021echo, wang2020public}. To curb this contagion, platforms implement a series of top-down interventions \cite{chandrasekharan2017you, wang2020efficient, mccabe2024post}. A common approach is to dissolve high-risk groups or block malicious accounts, aiming to sever the transmission chains of harmful information \cite{johnson2019hidden, thomas2023disrupting, mccabe2024post, yang2023rumor, wang2020efficient}. Accordingly, understanding how the interventions reshape or curb harmful information contagion dynamics has attracted considerable attention \cite{mekacher2023systemic}.

For decades, information spreading has been primarily modeled as compartmental frameworks on pairwise networks, with corresponding interventions typically formulated as node immunization or edge removal \cite{pastor2015epidemic, matamalas2018effective, liu2024efficient, han2026immunization}. Recent studies, however, have shown that real-world social contagion often involves higher-order interactions, such as group pressure and collective reinforcement, in which influence arises from non-additive group effects \cite{iacopini2019simplicial, de2020social, ferraz2023multistability, ferraz2024contagion, hao2025unified, oliveira2026rumor, mancastroppa2026higher}. Such interactions are fundamentally different from pairwise contacts. Specifically, they cannot be reduced to combinations of dyadic links but must be captured by higher-order networks, including hypergraphs and simplicial complexes \cite{boccaletti2023structure, pister2024stochastic, di2024percolation, hao2026identifying, malizia2025hyperedge, battiston2026collective}. Moreover, the resulting higher-order contagion can produce rich dynamical behaviors, such as bistability and explosive transitions \cite{st2021universal, han2024probabilistic, burgio2024triadic, malizia2025disentangling, zhao2025higher}. In this context, dissolving groups naturally corresponds to hyperedge removal, which has been shown to effectively mitigate contagion dynamics under the assumption of a static topology \cite{jhun2021effective}.

However, such static-topology assumption does not fully capture a key feature of real-world social systems: users can adapt to platform interventions \cite{johnson2019hidden, mekacher2023systemic}. Specifically, empirical evidence has shown that when a violating group is dissolved, its members may relocate to existing groups or establish new ones, leading to a bottom-up reorganization of the interaction topology \cite{horta2023deplatforming, monti2023online}. This adaptive reorganization can forge new pathways of transmission and thereby influence the effectiveness of interventions in curbing harmful information contagion. Yet how platform-induced interventions, such as group dissolution, compete with user adaptation remains largely unexplored.  

In this work, we propose an adaptive higher-order information contagion model by incorporating both platform-induced group dissolution and individual adaptive rewiring. Specifically, the platform intervention is governed by a baseline dissolution rate and a higher-order reinforcement parameter. Then, using hyperedge-based approximation, we develop a theoretical framework that shows good agreement with large-scale simulations. Notably, we observe complex impacts of group dissolution in adaptive topology. First, we identify a critical infection rate, above which increasing the dissolution rate counterintuitively exacerbates harmful information diffusion. Below this limit, increasing the dissolution rate induces a non-monotonic trend, initially amplifying the contagion before eradicating it via a discontinuous transition at a dissolution threshold. These findings reveal that group dissolution can successfully eradicate harmful information only within a specific infection-rate window, outside of which it may backfire and amplify harmful information. Furthermore, we show that higher-order reinforcement expands this infection-rate window, whereas rewiring homophily narrows it and raises the critical dissolution threshold, thereby weakening the efficacy of group dissolution. Moreover, these results are validated on real hypergraphs. Our work reveals the competition between platform-induced group dissolution and individual adaptive rewiring, providing key insights of curbing harmful information.

\section{Model and theoretical framework}
\subsection{Adaptive information contagion model} \label{sec:model_dynamics}

\begin{figure*}[!t]
\centering
\includegraphics[width=0.75\textwidth]{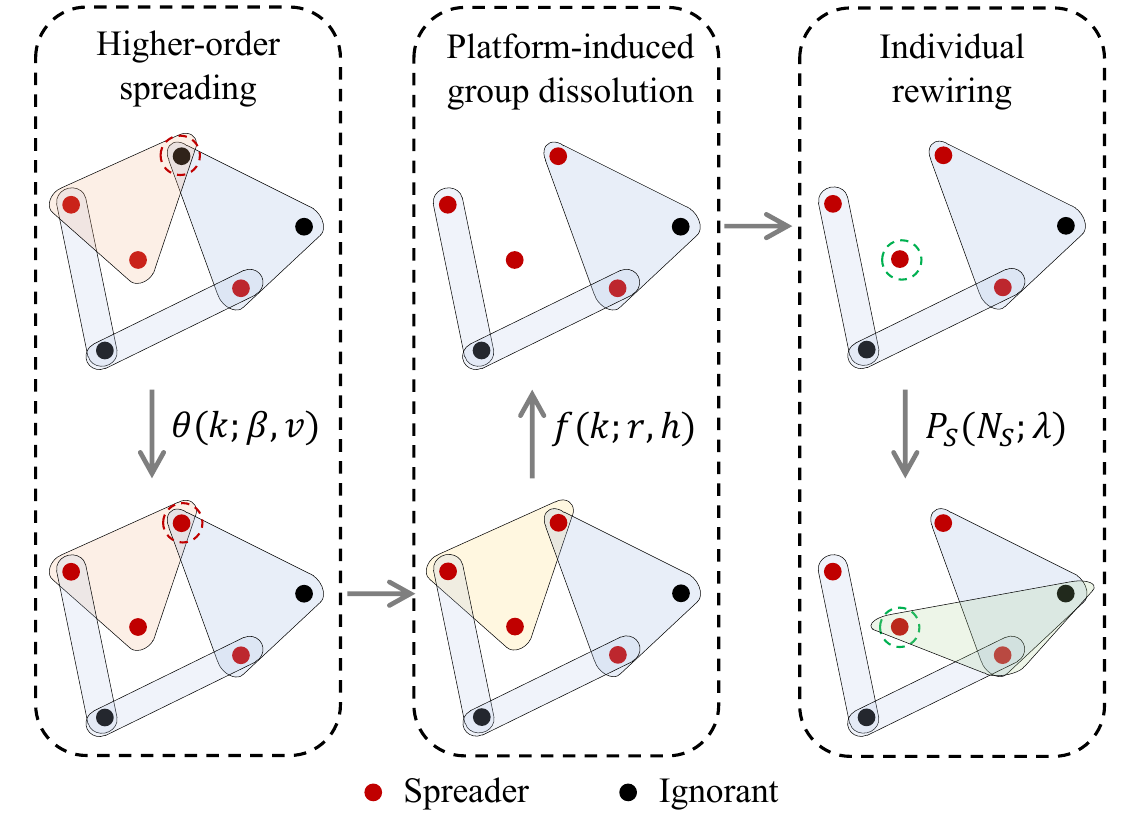}
\caption{Schematic illustration of the proposed model framework. This model depicts the coevolution of higher-order information contagion and structural changes. Specifically, the structural evolution is primarily driven by platform-induced group dissolution and individual adaptive rewiring, both of which depend on the dynamical state of nodes. In turn, these structural changes shape the subsequent contagion outcomes.}\label{fig1}
\end{figure*}

Our model captures the adaptive higher-order contagion of harmful information by incorporating platform-induced group dissolution and individual rewiring, as illustrated in Fig.~\ref{fig1}. We employ a hypergraph to represent the structured population of $N$ individuals, where nodes denote individuals and $m$-hyperedges encode groups involving $m$ members. For the contagion process, each individual alternates between two states: Spreader ($\mathrm{S}$) who actively disseminates the harmful information, and Ignorant ($\mathrm{I}$) who loses interest in or does not endorse it. Following the recent higher-order contagion mechanism \cite{st2021universal}, an ignorant individual in a hyperedge containing $k$ spreaders becomes infected at a rate $\theta_k=\beta k^v$, where $\beta$ is the baseline infection rate and $v$ represents nonlinear higher-order reinforcement. Moreover, each spreader becomes ignorant at a rate $\mu$. 

Crucially, we introduce adaptive structural evolution. First, social media platform often dissolves violating groups to curb the spread of harmful information \cite{johnson2019hidden, thomas2023disrupting}. This top-down intervention can be naturally modeled as hyperedge breaking mechanism. Specifically, we define the dissolution rate of a hyperedge containing $k\geq 1$ spreaders:
\begin{equation}
\pi_k=f(k,r,h),
\label{abs-eq}
\end{equation}
where $r$ is the baseline dissolution rate, and the parameter $h$ governs the sensitivity of this rate to the number of spreaders. This function must satisfy two conditions: (1) $f(1,r,h) = r$, and (2) $f(k-1,r,h)\leq f(k,r,h)$. The latter reflects the intuition that the likelihood of group dissolution increases with the number of spreaders. Without loss of generality, we instantiate this function as:
\begin{equation}
\pi_k=rh^{k-1},
\label{eq:disruption_rate}
\end{equation}
where $h\geq1$.

Moreover, individuals exhibit adaptive rewiring characteristics in the real world. For instance, spreaders from a dissolved group can reorganize into a new group to maintain their social activity \cite{monti2023online, horta2023deplatforming}. Here, to prevent network fragmentation and ensure analytical tractability, we assume that if an $m$-hyperedge is broken, a spreader from that group will select $m-1$ other nodes to form a new $m$-hyperedge. Specifically, the probability of sampling a spreader node is 
\begin{equation}
p_{\mathrm{S}}(t)=
\frac{\lambda N_{\mathrm{S}}(t)}
{\lambda N_{\mathrm{S}}(t)+N_{\mathrm{I}}(t)},
\label{eq:spreader_selection}
\end{equation}
where $N_{\mathrm{I}}(t)$ and $N_{\mathrm{S}}(t)$ denote the numbers of ignorant individuals and spreaders, respectively. The parameter $\lambda$ measures the rewiring homophily. Specifically, $\lambda=1$ represents random rewiring. $\lambda>1$ signifies a homophilic tendency, meaning that the reorganizing spreader is more likely to choose other spreaders, whereas $0<\lambda<1$ implies a heterophilic preference.

\subsection{Theoretical framework}
We then derive the dynamical equations to describe this adaptive higher-order contagion process. Nevertheless, it is a challenging problem due to the intricate co-evolution of hyperedge topology and node states. Inspired by recent works \cite{burgio2025characteristic, liu2026inhibition}, we use hyperedge-based approximation to tackle this complexity, formulated as follows.

For a hypergraph, we denote $E_m$ as the number of $m$-hyperedges. Let $L_{m,k}(t)$ be the number of $m$-hyperedges containing $k$ spreaders at time $t$. Thus, we have $\sum_{k=0}^{m}L_{m,k}(t)\equiv E_m$. In this setting, the average rate that an ignorant individual becomes infected can be approximated by 
\begin{equation}
\Theta(t)= \sum_m\sum_{k=1}^{m} \frac{(m-k)L_{m,k}(t)}{N_{\mathrm{I}}(t)}\beta k^v,
\label{eq:mean_spreading_pressure}
\end{equation}
where the term $(m-k)L_{m,k}/N_{\mathrm{I}}$ is the mean number of $m$-hyperedges containing $k$ spreaders that are incident to an ignorant individual. Thus, the dynamical equation of $N_{\mathrm{S}}(t)$ satisfies
\begin{equation}
\frac{\mathrm{d}N_{\mathrm{S}}}{\mathrm{d}t}
=-\mu N_{\mathrm{S}}+N_{\mathrm{I}}\Theta
=-\mu N_{\mathrm{S}}
+\beta\sum_m\sum_{k=1}^{m}
(m-k)k^v L_{m,k}.
\label{eq:spreader_dynamics}
\end{equation}

Moreover, note that Eq.~\eqref{eq:spreader_dynamics} depends on the time-varying hyperedge variable $L_{m,k}(t)$, which should be tracked. The temporal evolution of $L_{m,k}(t)$ is driven by two mechanisms: (1) node state transitions induced by the contagion process, and (2) the dissolution and formation of hyperedges resulting from adaptive structural updates. Therefore, the equations governing $L_{m,k}(t)$ are given by
\begingroup
\small
\begin{align}
\frac{\mathrm{d}L_{m,k}}{\mathrm{d}t}
={}&-[\theta_k+\Theta](m-k)L_{m,k}\nonumber\\
&-\mu kL_{m,k}\nonumber\\
&+[\theta_{k-1}+\Theta](m-k+1)L_{m,k-1}
+\mu(k+1)L_{m,k+1}\nonumber\\
&+\binom{m-1}{k-1}p_{\mathrm{S}}^{\,k-1}
(1-p_{\mathrm{S}})^{m-k}
\sum_{j=1}^{m}\pi_jL_{m,j}\nonumber\\
&-\pi_kL_{m,k},\nonumber\\
&\qquad 1\leq k\leq m,
\label{eq:hyperedge_dynamics}
\end{align}
\endgroup
where the first two lines account for node state transitions and the remaining lines arise from adaptive structural updates. In addition, $L_{m,0}=E_m-\sum_{k=1}^{m}L_{m,k}$ and  $N_{\mathrm{I}}(t)=N-N_{\mathrm{S}}(t)$. Together, Eqs.~\eqref{eq:spreader_dynamics} and \eqref{eq:hyperedge_dynamics} form a closed autonomous system.

\section{Results}
\subsection{Results on synthetic hypergraphs}

This section primarily explores whether, and to what extent, group dissolution can mitigate information contagion in the presence of individual adaptive behaviors. Using Gillespie simulations and theoretical analysis \cite{gillespie1977exact}, we focus on a simple but representative situation, i.e., adaptive contagion on synthetic $3$-uniform hypergraphs. Let $E$ denote the number of hyperedges. Then, $d=3E/N$ is the mean hyperdegree. In this case, the system has a spreader-free equilibrium at $(N_{\mathrm{S}}^*,L_{3,1}^*,L_{3,2}^*,L_{3,3}^*)=(0,0,0,0)$. The stability of this equilibrium determines whether harmful information can invade the system from an infinitesimal seed. By linearizing Eqs.~\eqref{eq:spreader_dynamics}-\eqref{eq:hyperedge_dynamics} around this equilibrium (see Appendix~\ref{sec:methods_threshold} for details), we can obtain the invasion threshold
\begin{equation}
\beta_c=
\frac{-C_1+\sqrt{C_1^2+4C_2C_0}}{2C_2},
\label{eq:invasion_threshold}
\end{equation}
where
\begin{align}
C_2&=2^{v+1}d(3\mu+2rh^2),\nonumber\\
C_1&=2d(3\mu+rh^2)(2\mu+rh)-2^v\mu rh^2,\nonumber\\
C_0&=\mu(3\mu+rh^2)(2\mu+rh).
\label{eq:threshold_coefficients}
\end{align}

\begin{figure*}[!t]
\centering
\includegraphics[width=0.8\textwidth]{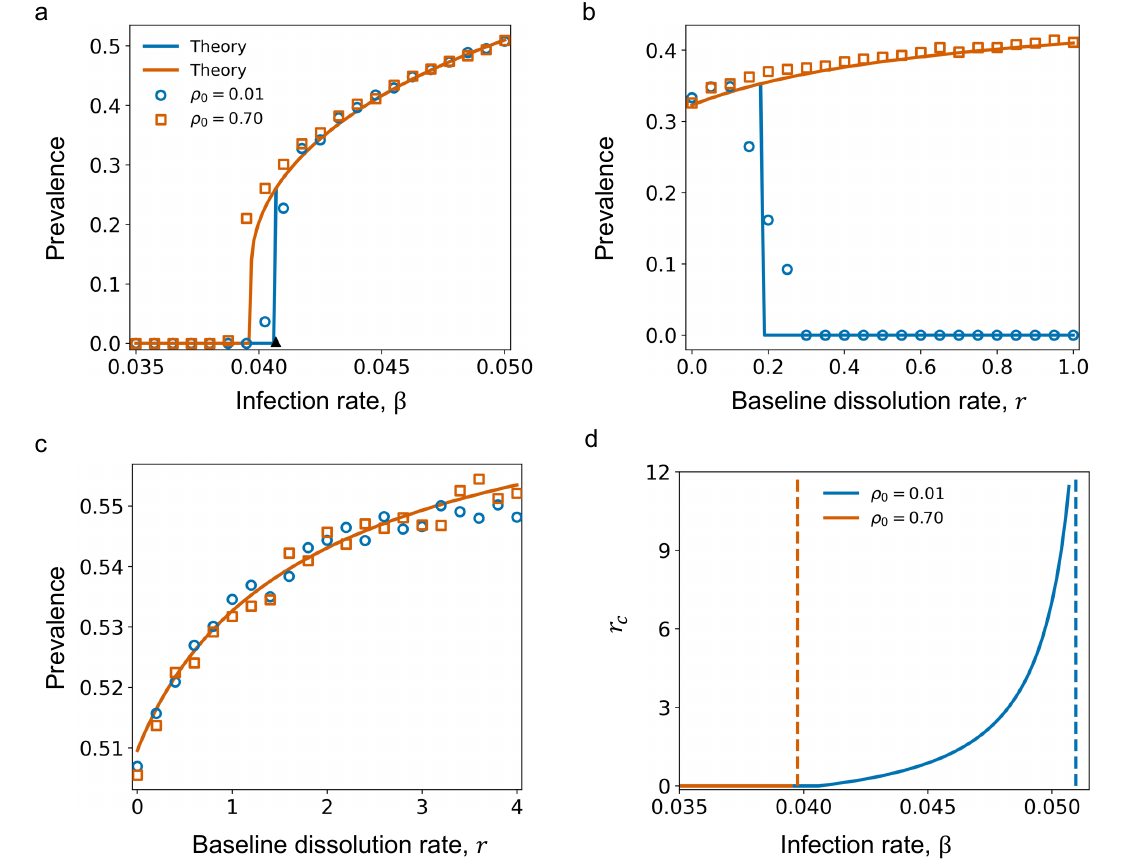}
\caption{Impact of baseline dissolution rate $r$. (a) Information prevalence is plotted against infection rate $\beta$ under different initial densities $\rho_0$, where $r=0.01$. Both theoretical lines predicted by Eqs.~\eqref{eq:spreader_dynamics}-\eqref{eq:hyperedge_dynamics} and invasion threshold (black triangle) calculated by Eq.~\eqref{eq:invasion_threshold} agree well with the symbols obtained from simulations. We then consider two representative cases: (b) $\beta=0.042$ which is close to the invasion threshold, and (c) $\beta=0.05$ which is well above it. For low $\rho_0$ in panel (b), increasing dissolution rate $r$ initially raises prevalence, but prevalence abruptly collapse to zero beyond a critical value $r_c$. However, in panel (c), increasing $r$ backfires and monotonically increases prevalence. Panel (d) plots $r_c$ versus $\beta$ for different $\rho_0$. The vertical asymptotes (dashed lines) bound the infection-rate window within which sufficiently large $r$ can eradicate information. Parameters: $N=10000$, $E=16000$, $v=2.8$, $\mu=0.5$, $h=1.0$, $\lambda=1.0$.}\label{fig2}
\end{figure*}

First, we explore the impact of the baseline dissolution rate $r$ under random rewiring ($\lambda=1$) and without higher-order reinforcement ($h=1$). Figure~\ref{fig2}(a) shows information prevalence against the infection rate under different initial conditions $\rho_0$. The good agreement between the analytical predictions and simulations validates the accuracy of our theoretical framework. We also find discontinuous phase transitions and bistability induced by the higher-order contagion mechanism, confirming classic findings \cite{wang2024epidemic}. We then examine how dissolution rate $r$ shapes contagion outcomes in two representative cases: $\beta=0.042$ (near the invasion threshold) and $\beta=0.05$ (well above it). For low $\rho_0$ in the former case (see Fig.~\ref{fig2}(b)), increasing $r$ initially promotes contagion, whereas prevalence abruptly collapses to zero via a discontinuous transition once $r$ exceeds a critical value $r_c$. By contrast, for $\beta=0.05$ (see Fig.~\ref{fig2}(c)), increasing $r$ only backfires, monotonically increasing prevalence without any abrupt collapse. This finding suggests that increasing $r$ is not universally effective, raising a key question: under what conditions can it eradicate harmful information contagion? Figure~\ref{fig2}(d) plots $r_c$ as a function of $\beta$ for different $\rho_0$. Notably, vertical asymptotes emerge in the $r_c-\beta$ curve. Each asymptote represents a critical infection rate: below this threshold, increasing $r$ beyond $r_c$ can eradicate harmful information, whereas above it, even the limit $r\rightarrow \infty$ fails to do so. Moreover, large $\rho_0$ shifts the asymptote toward smaller $\beta$, i.e., shrinking this infection-rate window where dissolution mechanism remains effective.

\begin{figure*}[!t]
\centering
\includegraphics[width=0.8\textwidth]{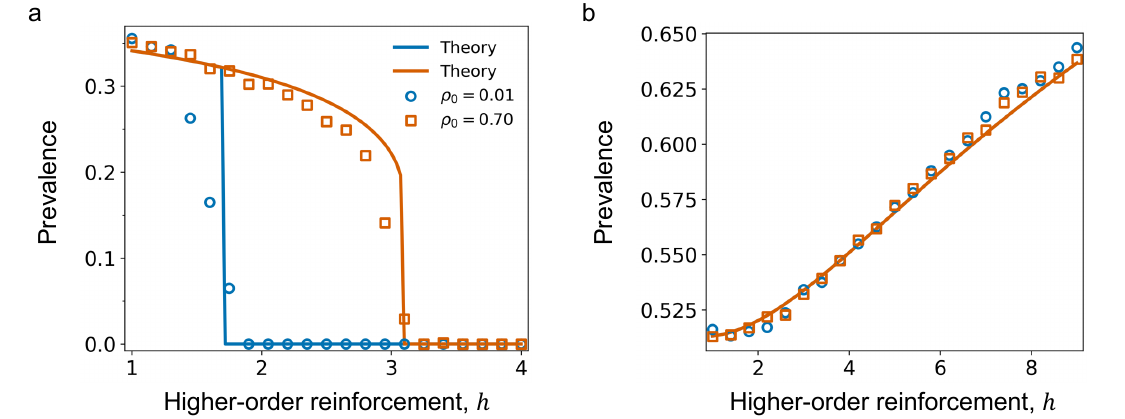}
\caption{Impact of higher-order reinforcement parameter $h$. Information prevalence is plotted against $h$ under (a) $\beta=0.042$ (near the invasion threshold) and (b) $\beta=0.05$ (well above it). The lines are theoretical predictions from Eqs.~\eqref{eq:spreader_dynamics}-\eqref{eq:hyperedge_dynamics}, while symbols correspond to simulation results. In panel (a), as $h$ increases, prevalence initially decreases and then abruptly drops to zero via a discontinuous transition once $h$ exceeds a critical value. In contrast, panel (b) shows that increasing $h$ promotes contagion. Parameters: $N=10000$, $E=16000$, $v=2.8$, $\mu=0.5$, $r=0.1$, $\lambda=1.0$.}\label{fig3}
\end{figure*}

Figure~\ref{fig3} further examines how higher-order reinforcement parameter $h$ shapes information contagion under the above representative values of $\beta$. Figure~\ref{fig3}(a) shows the case $\beta=0.042$. As $h$ increases, the prevalence monotonically decreases and undergoes a discontinuous phase transition at a critical value $h_c$, where it drops abruptly from a finite value to zero. Moreover, we observe bistability phenomenon, which reveals that a large initial density $\rho_0$ substantially increases the threshold $h_c$. Conversely, the scenario with $\beta=0.05$ exhibits markedly different results, as shown in Fig.~\ref{fig3}(b). Specifically, the prevalence increases monotonically with $h$. These findings suggest that the efficacy of higher-order reinforcement in suppressing harmful information highly depends on the infection rate $\beta$: increasing $h$ can eradicate contagion for small $\beta$, but yields adverse effects for large $\beta$.

\begin{figure*}[!t]
\centering
\includegraphics[width=0.8\textwidth]{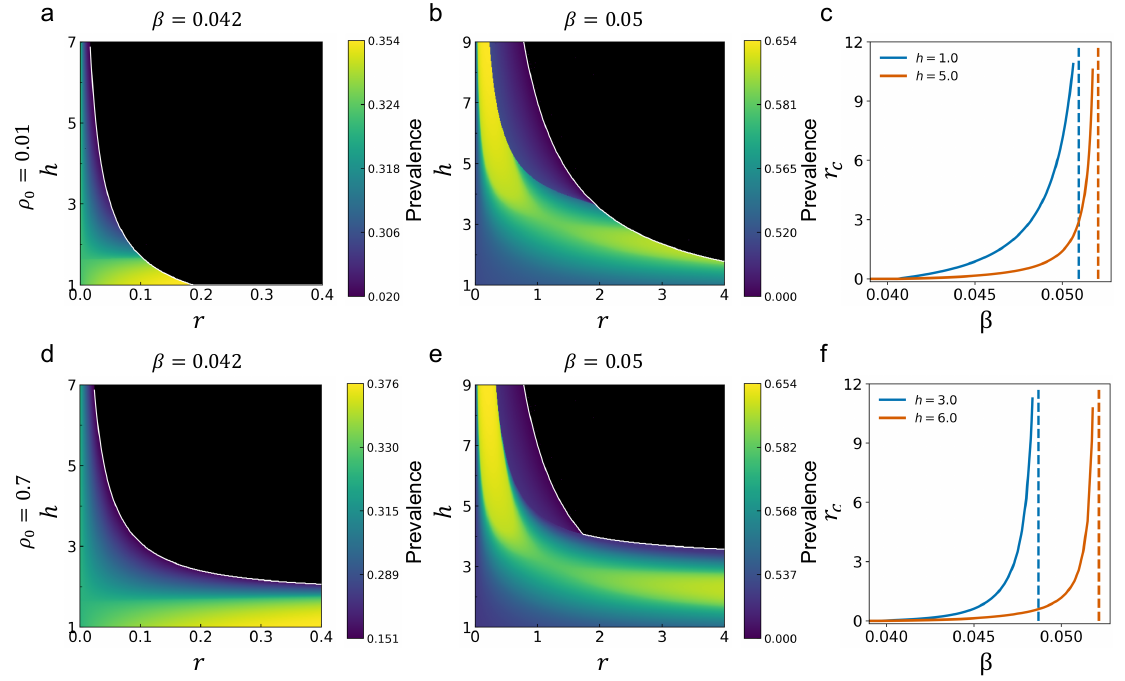}
\caption{Synergistic impact of baseline dissolution rate $r$ and reinforcement parameter $h$. Phase diagrams for information prevalence are plotted to show the joint effects of $r$ and $h$ under (a) $\beta=0.042$ and (b) $\beta=0.05$. The white lines indicate the critical values above which harmful information is eradicated. We find that a larger $h$ can substantially decrease the dissolution threshold $r_c$. Panel (c) shows $r_c$ versus infection rate $\beta$ for different $h$. Notably, increasing $h$ shifts the vertical asymptote (dashed line) toward larger $\beta$, thereby expanding the infection-rate window in which group dissolution can eradicate contagion. Compared to panels (a)-(c), panels (d)-(f) exhibit similar synergistic impact of $r$ and $h$ in the high-$\rho_0$ case. Parameters: $v=2.8$, $\mu=0.5$, $\lambda=1.0$.}\label{fig4}
\end{figure*}

Furthermore, we present phase diagrams to illustrate the joint impact of the dissolution rate $r$ and reinforcement parameter $h$, which together characterize platform-induced group dissolution. Figures~\ref{fig4}(a)-(c) depict the scenario with a low initial density $\rho_0$. Across various infection rates $\beta$, the dissolution threshold $r_c$ required to eradicate harmful information decreases with $h$, as indicated by the white lines in Figs.~\ref{fig4}(a) and (b). Moreover, Fig.~\ref{fig4}(c) presents $r_c$ as a function of $\beta$ for different values of $h$. The curves show that $r_c$ increases with $\beta$ and approaches a vertical asymptote. This asymptote defines the upper bound of the infection-rate window within which information can be eradicated by increasing $r$ above $r_c$. Beyond this bound, eradication becomes impossible even in the limit $r\rightarrow\infty$. Notably, larger $h$ shifts the asymptote towards higher $\beta$, demonstrating that higher-order reinforcement expands the infection-rate window over which group dissolution remains effective. Moreover, Figures~\ref{fig4}(d)-(f) show a qualitatively similar synergistic pattern of $r$ and $h$ in the high-$\rho_0$ case: increasing $h$ not only lowers $r_c$, but also broadens the effective infection-rate range for the dissolution mechanism.

\begin{figure*}[!t]
\centering
\includegraphics[width=0.8\textwidth]{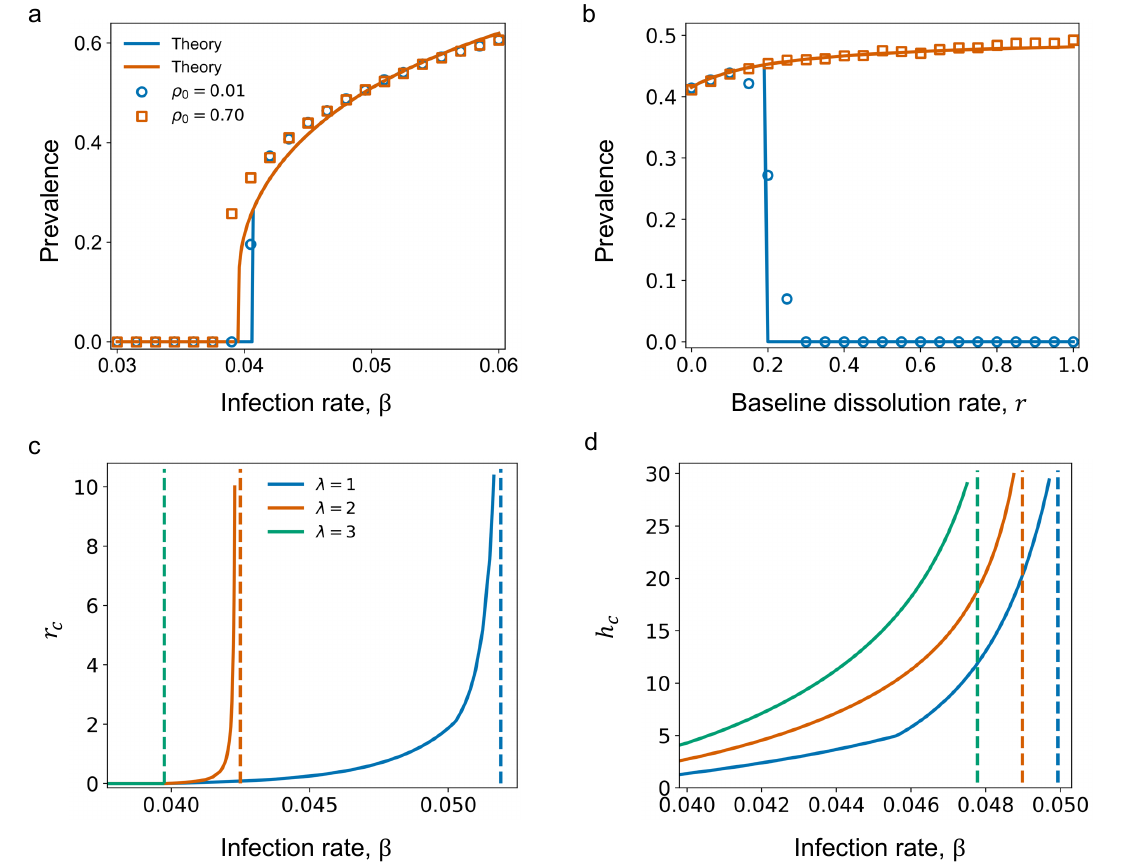}
\caption{How rewiring homophily $\lambda$ shapes the impact of group dissolution mechanism. (a) Information prevalence is plotted for the homophilic rewiring parameter $\lambda=2.0$. The theoretical predictions (lines) are computed by Eqs.~\eqref{eq:spreader_dynamics}-\eqref{eq:hyperedge_dynamics}, which agree well with simulations (symbols). (b) In this setting, we select $\beta=0.045$ (near the invasion threshold), and plot information prevalence versus baseline dissolution rate $r$. As $r$ increases, prevalence for low $\rho_0$ initially rises before abruptly dropping to zero via a discontinuous transition at a dissolution threshold $r_c$, whereas prevalence monotonically increases for high $\rho_0$. (c) The dissolution threshold $r_c$ is plotted for different $\lambda$, with $\rho_0=0.7$ and $h=4.0$. We find that rewiring homophily $\lambda$ substantially increases $r_c$ and shifts the vertical asymptote (dashed line) to smaller $\beta$, both of which diminish the dissolution mechanism's efficacy in suppressing information. Panel (d) shows the reinforcement threshold $h_c$ required to eliminate information under different $\lambda$, where $\rho_0=0.7$ and $r=0.2$. The results show that $\lambda$ significantly increases $h_c$ and move the asymptote towards smaller $\beta$. Parameters: $N=10000$, $E=16000$, $v=2.8$, $\mu=0.5$.}\label{fig5}
\end{figure*}

Figure~\ref{fig5} further explores how individual rewiring homophily $\lambda$ influences the efficacy of platform-induced group dissolution mechanism. Fig.~\ref{fig5}(a) presents the prevalence as a function of $\beta$ under homophilic rewiring $\lambda=2$. The theoretical results agree well with the simulations, confirming the validity of our theoretical framework under non-random rewiring. In homophilic setting, Fig.~\ref{fig5}(b) examines how the baseline dissolution rate $r$ shapes the contagion outcomes. For low $\rho_0$, information prevalence initially rises with $r$ but then collapses to zero through a discontinuous transition at the dissolution threshold $r_c$. In contrast, for high $\rho_0$, the prevalence increases monotonically with $r$, without abrupt transition to spreader-free state. These findings are qualitatively consistent with those obtained under random rewiring, indicating two key insights. First, the dissolution mechanism operates effectively only within a bounded regime, beyond which increasing $r$ may backfire and instead facilitate the spread of harmful information. Second, within this regime, $r$ must exceed a critical threshold $r_c$ to successfully eradicate contagion. Furthermore, Fig.~\ref{fig5}(c) plots $r_c$ against the infection rate $\beta$ under different rewiring homophily $\lambda$. Results show that increasing $\lambda$ not only raises $r_c$ but also shifts the asymptote towards smaller $\beta$, thereby narrowing the infection-rate regime in which the dissolution mechanism remains effective. Fig.~\ref{fig5}(d) illustrates the critical higher-order reinforcement $h_c$ versus $\beta$ under different $\lambda$. Similarly, as $\lambda$ grows, $h_c$ rises and the vertical asymptote moves toward smaller $\beta$. Overall, rewiring homophily undermines the efficacy of platform-induced group dissolution in suppressing harmful information.

\subsection{Results on real hypergraphs}

Here, we examine the adaptive model on an empirical hypergraph constructed from the SocioPatterns high school dataset \cite{mastrandrea2015contact, genois2018can}. The construction procedure is as follows. We first aggregate the interactions over 5-minute intervals. To reduce stochastic fluctuations caused by finite-size effects, we then enlarge the aggregated hypergraph by 20 times \cite{young2017construction} (see Appendix~\ref{sec:real_hyperedge} for details). The resulting hypergraph includes 6484 nodes, 5568 2-hyperedges, and 9680 3-hyperedges.

\begin{figure*}[!t]
\centering
\includegraphics[width=0.8\textwidth]{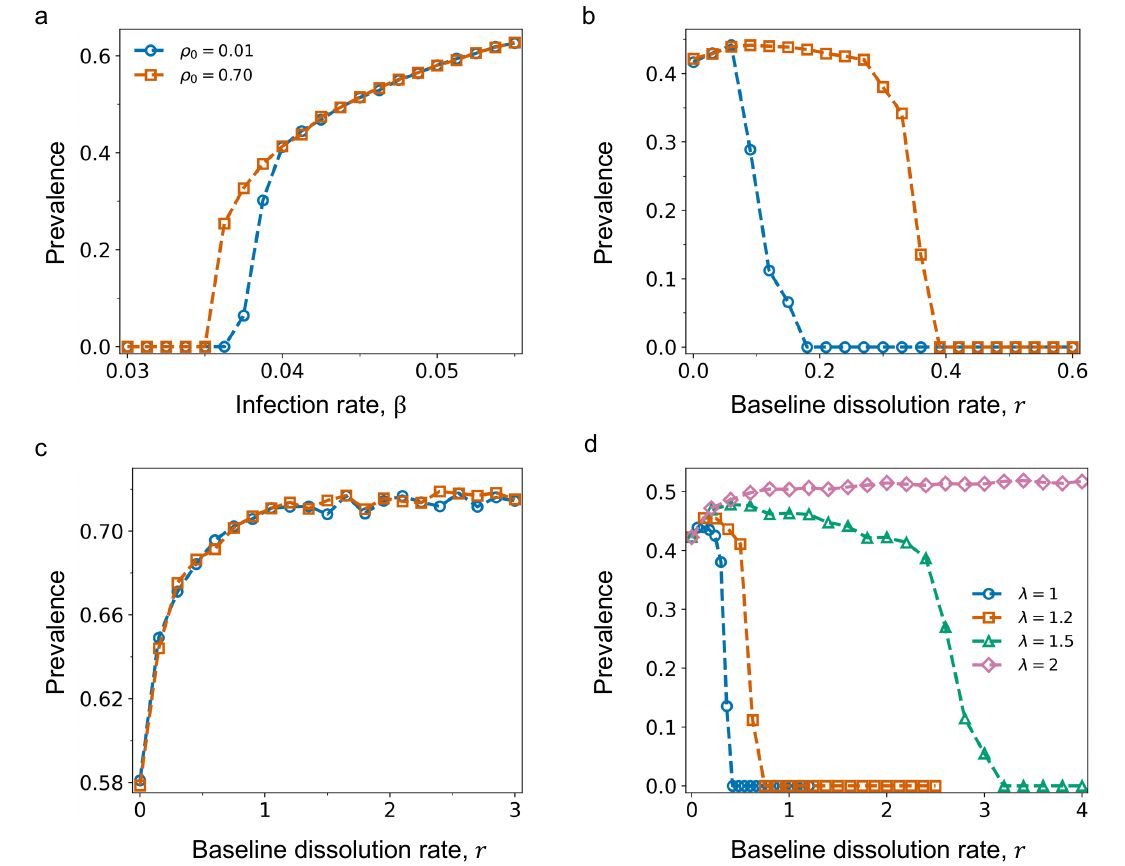}
\caption{Simulation results on a real hypergraph.
(a) Information prevalence as a function of $\beta$. The parameters are set to $r=0.01$, $h=1.0$, $\lambda=1.0$. Panels (b) and (c) further plot the prevalence against baseline dissolution rate with $h=4$ under two scenarios: $\beta=0.041$ (close to the invasion threshold) and $\beta=0.05$ (well above it), respectively. As $r$ increases, the prevalence for $\beta=0.041$ initially rises before abruptly dropping to zero at a critical threshold $r_c$, whereas it increases monotonically for $\beta=0.05$. (d) Prevalence as a function of $r$ for different $\lambda$ when $\beta=0.041$ and $\rho_0=0.7$. We find that $r_c$ increases with $\lambda$. Parameters: $v=3.0$, $\mu=0.5$.}\label{fig6}
\end{figure*}

Figure~\ref{fig6}(a) shows information prevalence as a function of the infection rate $\beta$ on this empirical hypergraph, revealing the emergence of bistability. We therefore choose two representative infection rates to examine the effect of platform-induced group dissolution: one near the invasion threshold and the other well above it. The results reproduce the qualitative behaviors observed on synthetic hypergraphs. At the lower infection rate (see Fig.~\ref{fig6}(b)), increasing dissolution rate initially enhances the prevalence, which is followed by an abrupt collapse to zero at a critical dissolution rate $r_c$. In contrast, at the higher infection rate (see Fig.~\ref{fig6}(c)), prevalence increases monotonically with the dissolution rate. These results indicate that the eradication effect of group dissolution is confined to an infection-rate window and requires the dissolution rate to exceed a critical threshold. Furthermore, Fig.~\ref{fig6}(d) shows that increasing rewiring homophily raises $r_c$ and may even prevent the system from reaching the spreader-free state, thereby weakening the ability of group dissolution in suppressing contagion. Taken together, these results on empirical hypergraph confirm the robustness of our findings.

\section{Conclusion}\label{sec4}

 This work proposes an adaptive higher-order contagion model for harmful information that captures the interplay between platform-induced group dissolution and individual adaptive rewiring. We develop a theoretical framework that accurately predicts the resulting coevolutionary dynamics. Notably, in contrast to the efficacy of group dissolution under static topological assumptions \cite{jhun2021effective}, our work reveals complex effects of such platform intervention on suppressing information contagion when confronted with adaptive human behaviors.

First, increasing the dissolution rate does not necessarily suppress information contagion but may instead become counterproductive under certain conditions, thereby exacerbating its spread. Specifically, there exists an upper bound of effective infection-rate window, beyond which the dissolution mechanism backfires and unexpectedly promotes contagion. Within this window, increasing the dissolution rate initially raises prevalence, but then triggers an abrupt collapse to zero at a dissolution threshold, signaling a discontinuous phase transition. These findings underscore two prerequisites for eradicating harmful information: the infection rate is low to keep the system within the effective window, and the dissolution rate surpasses the critical dissolution threshold. In addition, a high initial prevalence can further tighten these prerequisites, making eradication more difficult.

Furthermore, we demonstrate that higher-order reinforcement significantly enhances the ability of the dissolution mechanism in suppressing contagion by expanding its effective window and lowering the dissolution threshold. We also investigate the role of individual rewiring homophily. Our results show that homophilic rewiring narrows the infection-rate window where the dissolution mechanism remains effective, weakening the efficacy of this platform intervention. Similar results are obtained from simulations on an empirical hypergraph, demonstrating the robustness of our findings.

Our work systematically explores how platform-induced group dissolution interacts with adaptive human behaviors in harmful information contagion, inspiring the reconsideration of effective control strategies. In addition, while our current analysis is grounded in a classic contagion model, the underlying mechanism of adaptive structural coevolution can be naturally generalized to other complex dynamical models \cite{liu2026emergent, alvarez2021evolutionary, jiang2026nonlinear, lamata2025hyperedge, xiang2026controllability}. Moreover, exploring diverse adaptive rewiring rules and extending this framework to multilayer networks to capture cross-platform migration warrant further investigation \cite{chang2026oscillatory, pal2024global, wang2020epidemic}.  

\appendices

\section{Invasion threshold on 3-uniform hypergraphs}\label{sec:methods_threshold}

At the spreader-free equilibrium
$(N_{\mathrm{S}}^*,L_{3,1}^*,L_{3,2}^*,L_{3,3}^*)=(0,0,0,0)$, we linearize the system depicted by Eqs.~\eqref{eq:spreader_dynamics}-\eqref{eq:hyperedge_dynamics} and obtain the Jacobian matrix
\begingroup
\setlength{\arraycolsep}{2pt}
\begin{equation}
J_*=
\begin{pmatrix}
-\mu & 2\beta & 2^v\beta & 0\\
0 & 2(d-1)\beta-\mu & 2^vd\beta+2\mu+rh & rh^2\\
0 & 2\beta & -2^v\beta-2\mu-rh & 3\mu\\
0 & 0 & 2^v\beta & -3\mu-rh^2
\end{pmatrix},
\label{eq:methods_jacobian}
\end{equation}
\endgroup
where $d=3E/N$ represents the mean hyperdegree. Its block-upper-triangular form gives one eigenvalue $-\mu<0$. The remaining three eigenvalues are determined by the lower-right $3\times 3$ block of matrix, denoted by $A_3(\beta)$. At $\beta=0$, all eigenvalues are negative. For $\beta>0$, $A_3(\beta)$ is an irreducible Metzler matrix, so its dominant eigenvalue is real. Therefore, loss of stability can occur only when $\det A_3(\beta)=0$.

$\det A_3(\beta)$ satisfies
\begin{align}
\det A_3(\beta)
={}&[2(d-1)\beta-\mu]\nonumber\\
&\quad\times\bigl[(3\mu+rh^2)(2\mu+rh)
+2^vrh^2\beta\bigr]\nonumber\\
&+2\beta(3\mu+rh^2)
\bigl[2^vd\beta+2\mu+rh\bigr]\nonumber\\
&+2^{v+1}rh^2\beta^2\nonumber\\
={}&C_2\beta^2+C_1\beta-C_0,
\label{eq:methods_threshold_polynomial}
\end{align}
where $C_0$, $C_1$, and $C_2$ are given in Eq.~\eqref{eq:threshold_coefficients}. Since $C_2>0$ and $C_0>0$, the two roots have product $-C_0/C_2<0$ and therefore have opposite signs. The unique positive root is
\begin{equation}
\beta_c=
\frac{-C_1+\sqrt{C_1^2+4C_2C_0}}{2C_2},
\label{eq:methods_threshold_root}
\end{equation}
This is the invasion threshold for an infinitesimal spreader seed.

\section{Construction of the real hypergraph}\label{sec:real_hyperedge}

We construct the real hypergraph from the SocioPatterns high-school dataset, which records face-to-face contacts between students at a temporal resolution of 20 seconds \cite{mastrandrea2015contact,genois2018can}. The contacts are first aggregated within non-overlapping 5-minute windows. For each aggregated snapshot, maximal cliques are identified and interpreted as simultaneous group interactions. We retain only interactions represented by 2-hyperedges and 3-hyperedges. Cliques involving more than three individuals are decomposed into all constituent 3-hyperedges. The resulting hyperedges are then ranked according to their occurrence frequencies, and we keep the top 10\% to construct the empirical hypergraph. To reduce stochastic fluctuations caused by finite-size effect, we augment the hypergraph by 20 times using the algorithm described in Ref.~\cite{young2017construction}. The final hypergraph contains 6484 nodes, 5568 2-hyperedges, and 9680 3-hyperedges.

\bibliographystyle{IEEEtran}
\bibliography{adaptive-information}

\end{document}